\documentclass[aps,twocolumn,showpacs]{revtex4}
\usepackage{epsfig}
\usepackage{amsbsy}
\usepackage{amsmath}
\usepackage{graphicx}
\usepackage{latexsym}

\begin{document}

\title{Heat conduction in one-dimensional Yukawa chains}
\author{Bambi Hu$^{1,2}$, Haibin Li$^{1,3}$, and Bai-Song Xie$^{1,4}$}
\altaffiliation{Author to whom correspondence should be addressed.}
\affiliation{$^1$Department of Physics and Centre for Nonlinear Studies, Hong Kong
Baptist University, Hong Kong, China\\
$^2$Department of Physics, University of Houston, Houston Tx 77204-5506\\
$^3$Zhejiang Institute of Modern Physics, Zhejiang University, Hangzhou
310027, China\\
$^4$Institute of Low Energy Nuclear Physics, Beijing Normal University,
Beijing 100875, China}
\date{\today}

\begin{abstract}
Heat conduction in one-dimensional Yukawa chains is investigated. It is
shown numerically that it has the abnormal heat conduction which is
proportional to the system size. Effects of asymmetric external potential,
the modified Frenkel-Kontorova one, on the heat conduction of system are
also studied. It is found that the asymmetric property of external potential
can induce asymmetric thermal conductivity that can be used to be a
effective thermal rectifier. In certain of system parameters the heat flux
are significantly different for two opposite direction. One can
parametrically control the heat flux through this system by changing the
potential strength and width.
\end{abstract}

\maketitle

Recent years many interesting works have been addressed to the problem of
heat conduction \cite{lep1,lep2,hu1,cas1,pos1,hu2,sav1,alo1,cas2}. It is
well known that heat conductivity of one dimensional (1D) lattice is an
important problem which is related to microscopic foundation of the Fourier
law. It also provides an abundant example for the microscopic origin of the
macroscopic irreversibility phenomenon. Detailed review of the problem is
presented in the recent paper \cite{lep1}.

Except of extensive numerical studies focusing on the validity of Fourier
law $J=-\kappa dT/dx$ for many models such as the Fermi-Pasta-Ulam (FPU)
model \cite{lep2,hu1}, the ding-a-ling model \cite{cas1,pos1}, the
Frenkel-Kontorova (FK) model \cite{hu2,sav1} and the Lorenz-gas model \cite%
{alo1} and so on, some other different and important problems have also
attracted people attention. For example most recently Casati \textit{et al.} %
\cite{cas2} investigated the possibility to control the energy transport
inside a nonlinear 1D chain connecting with two thermostats at different
temperatures. They emphasized that controlling heat conduction by
nonlinearity opens a new possibility to design a thermal rectifier, i.e., a
lattice that carries heat preferentially in one direction.

In this Letter we investigate heat conduction problem for a new type of
model wherein there is a Yukawa-Coulomb interaction potential. It arises
from dust-grain plasma which is extensively studied recently due to its new
features with dust crystal or collective waves \cite%
{lin1,tho1,vla1,xie1,xie2,sam1}. Our study shows that the heat conductivity
is abnormal in absence of external field as $\kappa \propto N$ as in
harmonic chains. Under FK\ field, however, it is normal which is similar to
the cases of former studies but with harmonic or/and FPU interactions \cite%
{hu2}. Surprisingly for the dust chains with real wake potential \cite{vla1}%
, modified FK field, the heat conduction exhibits a mixture property of
normal and abnormal behaviors. It is found that the asymmetric property of
external wake potential can induce asymmetric thermal conductivity that can
be used to be a thermal rectifier. In a certain of system parameters, by
changing the strength and width of wake potential, the heat fluxes for two
opposite direction can be different remarkably.

\begin{figure}[tbp]
\hskip -2cm \vskip -1cm \includegraphics[width=0.60\textwidth]{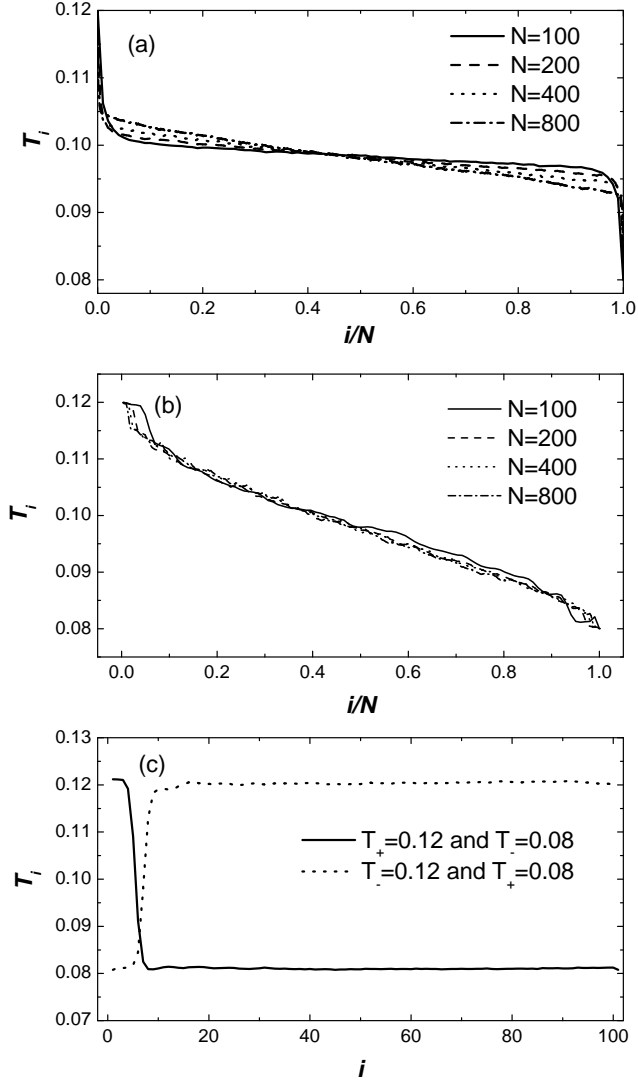} %
\vskip -0.5cm
\caption{Temperature profiles for different external potentials of
field-free (a), FK (b) and FK-modified (c). The imposed temperature are $%
T_{+}=0.12$ and $T_{-}=0.08$. The parameters are chosen as $\protect\gamma %
_{1}=1$ and $\protect\gamma _{2}=0.3$ (b), $2$ (c). Averages are carried
over a time interval $10^{7}$ by dropping a transient $10^{6}$.}
\end{figure}

In a general form the Hamiltonian of the considered system is written as%
\begin{equation}
H=\sum_{i}h_{i},\;\;\;\;h_{i}=\frac{p_{i}^{2}}{2}+V(x_{i-1,}x_{i})+U(x_{i})
\label{e1}
\end{equation}%
where $V(x_{i-1,}x_{i})$ stands for the interaction potential of the
nearest-neighbor particles and $U(x_{i})$ is a external on-site potential.
For dust particles system we have the interaction potential (refer to \cite%
{xie2}) 
\begin{equation}
V(r_{ij})=\frac{q^{2}}{r_{ij}}\exp (-\frac{r_{ij}}{\lambda _{d}})-\alpha 
\frac{q^{2}}{r_{ij}}  \label{e2}
\end{equation}%
where $\alpha $ is the parameter that ranges from $0$ to $1$, $\lambda _{d}$
is the plasma Debye length and $q$ is the dust grain charges, while $r_{ij}$
is the distance between particles $i$-th and $j$-th. For simplicity we have
assumed that the masses $m_{d}$ and charges of all the grains $q$ are the
same and constant, while for convenience we shall choose parameter $\alpha $
that makes the neighbor dust particle is located at the bottom of $V$ (see
more clear in the following). Note that the plasma electrons and ions do not
appear explicitly in the simulations but their effects, which result in an
attractive force to dust, are taken into account in the effective
interaction potential as the second term of Eq.(\ref{e2}). On the other hand
in dusty plasma we know that there exist wake potential acting the dusts
along downstream \cite{vla1} 
\begin{equation}
U(r_{ti})=\frac{qq_{t}}{r_{ti}}\frac{2}{1-M^{-2}}\cos (\frac{r_{ti}}{\lambda
_{d}\sqrt{M^{2}-1}})  \label{e4}
\end{equation}%
where $q_{t}$ is the upstream test dust particle charge, $M>1$ is the Mach
number of ion flow and $r_{ti}$ is the distance of $i$-th particle from the
test particle. The potential (\ref{e4}) is named as the FK-modified one in
this paper because it is different from FK potential with varying amplitude.

Let us turn to the side of the motion of system. A general approach is
employed here that are used by many authors like Lepri \textit{et al.} and
Hu \textit{et al.}, namely, two Nose-Hoover thermostats are put on the first
and last particle, keeping the temperature at $T_{+}$ and $T_{-}$,
respectively. For simplicity and convenience, firstly we need to normalize
the basic physical quantities: the mass by the particle mass $m_{d}$, the
space distance by the dust inter-particle distance $a_{d}$ and the time by
the $\sqrt{ma_{d}^{3}/q^{2}}$ respectively. Then the equations of motion for
particles are%
\begin{equation*}
\overset{\cdot \cdot }{x}_{1}=-\xi _{+}\overset{\cdot }{x}%
_{1}+f_{1}-f_{2},\;\;\xi _{+}=\frac{\overset{\cdot }{x}_{1}^{2}}{T_{+}}-1
\end{equation*}%
\begin{equation}
\overset{\cdot \cdot }{x}_{i}=\ f_{i}-f_{i+1},\;\;i=2,...,N-1  \label{e5}
\end{equation}%
\begin{equation*}
\overset{\cdot \cdot }{x}_{N}=-\xi _{-}\overset{\cdot }{x}%
_{N}+f_{N}-f_{N-1},\;\;\xi _{-}=\frac{\overset{\cdot }{x}_{N}^{2}}{T_{-}}-1
\end{equation*}

\begin{figure}[tbp]
\hskip -2cm \vskip -1cm \includegraphics[width=0.60\textwidth]{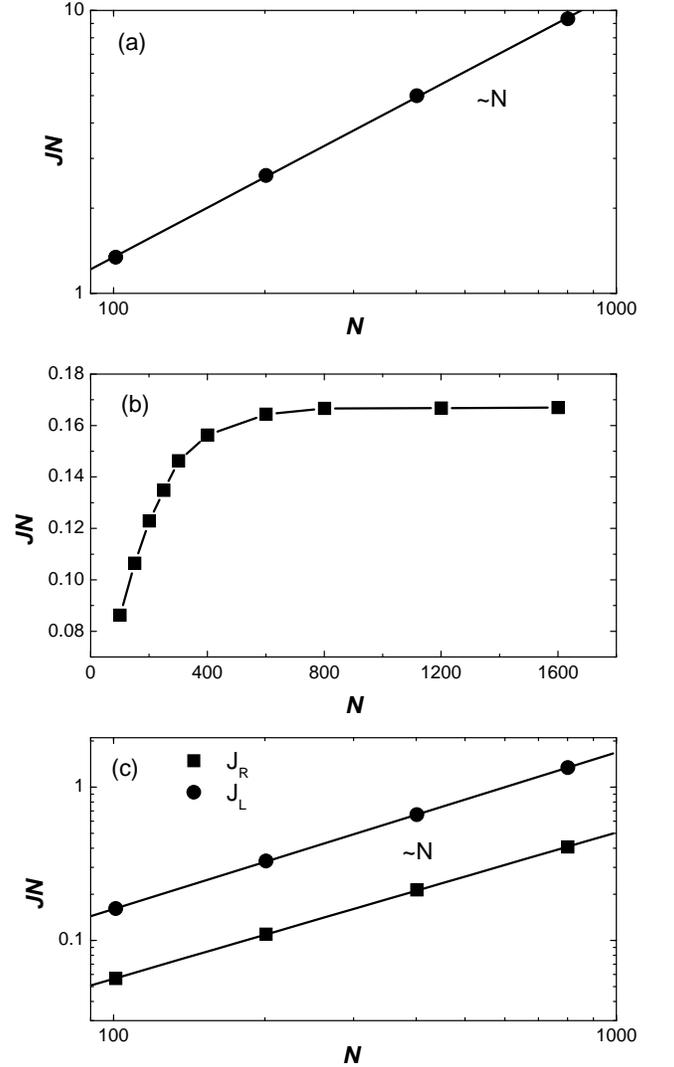} %
\vskip -0.5cm \vskip -2mm
\caption{Scaling of the heat flux $JN$ with the number of particles $N$ in
the Yukawa interaction chains for different external potentials of
field-free (a), FK (b) and FK-modified (c). System parameters and others are
same as in Fig.1}
\end{figure}

where $f_{i}=-V^{^{\prime }}(x_{i}-x_{i-1})-U^{^{\prime }}(x_{i})$ is the
force. Usually the fixed boundary conditions $x_{0}\equiv \mathrm{const}$
and $x_{N+1}\equiv \mathrm{const}$ are assumed. Now the interaction
potential in normalized form can be written 
\begin{equation}
V(x_{i-1,}x_{i})=\left( \exp [-\chi (x_{i}-x_{i-1})]-\alpha \right)
/(x_{i}-x_{i-1})  \label{e6}
\end{equation}%
where parameter $\chi =a_{d}/\lambda _{d}$ is the ratio of inter-particle
distance to the plasma Debye length. Many studies show that when $\chi $ is
large the nearest neighbor interaction is appropriate instead of
particle-pair interaction \cite{sam1}. In our following numerical studies $%
\chi =5$ is generally used. The particles are arranged initially at
equilibrium positions of $V$ and the normalized inter-distance equals to 
\textrm{1}, therefore, it means that parameter $\alpha $, adjusting the
attraction of plasma to dust particle, must satisfy $\alpha =(1+\chi )\exp
(-\chi )$. For the external FK-modified potential Eq.(\ref{e4}) we have its
normalized form 
\begin{equation}
U(x_{i})=\gamma _{2}\cos (\gamma _{1}x_{i})/x_{i},  \label{e7}
\end{equation}%
where $\gamma _{1}=\chi /\sqrt{M^{2}-1}$ and $\gamma _{2}=2\beta /(1-M^{-2})$
with $\beta =q_{t}/q$ the ratio of upstream test particle charge to
downstream particles charge. Now we denote $x_{i}$ as the position of $i$-th
dust particle in the chain from the origin of test dust particle $%
x_{0t}\equiv 0$. In correspondence for the normalized external FK potential
one has $U(x_{i})=\gamma _{2}\cos (\gamma _{1}x_{i})$. The different
temperature profiles for different external potentials of field-free, FK and
FK-modified are shown in Fig.1. The imposed temperature are $T_{+}=0.12$ and 
$T_{-}=0.08$ and $\gamma _{1}=1$. The potential strengths are $\gamma
_{2}=0.3$ and $2$ for FK and FK-modified, respectively. Obviously the
temperature gradients are constructed globally only in the FK case.

\begin{figure}[tbp]
\vskip -2cm \includegraphics[width=0.50\textwidth]{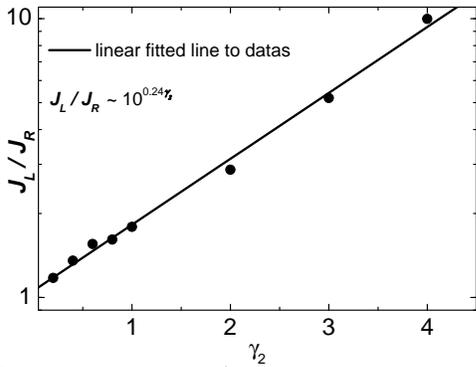} \vskip -7.5cm %
\vskip -2mm
\caption{Dependence of $J_{L}/J_{R}$, the heat flux ratio of opposite
direction for the Yukawa-Coulomb model, on parameter $\protect\gamma _{2}$,
the strength of the FK-modified on-site external potential. The imposed
temperature are $T_{+}=0.12$ and $T_{-}=0.08$ for $J_{R}$ and $T_{+}=0.08$
and $T_{-}=0.12$ for $J_{L}$. $\protect\gamma _{1}=1$ is given.}
\end{figure}

Usually the heat flux at the $i$-th position is given by $J_{i}(t)=\frac{1}{2%
}(\overset{\cdot }{x}_{i}+\overset{\cdot }{x}_{i+1})f_{i+1}$. Numerically
the time average $J=\langle J_{i}(t)\rangle $ is independent of the index
for long time enough. We plot in Fig.2 the heat flux $JN$ vs the system size 
$N$ for three different cases of field-free (a), FK (b) and FK-modified (c).
Obviously the abnormal heat conductivity $\kappa \propto N$ is obtained for
purely Yukawa interaction chains, which is similar to the case of harmonic
situation. In the Yukawa chains for standard FK external potential, although
it seems unclear about whether there exists phase transition of heat
conduction between normal and abnormal \cite{hu2,sav1}, our results
indicates that the normal heat conductivity exists, which support the
conclusion of normal heat conduction in Ref. \cite{hu2}. From Fig.2 (c),
however, we can see that it is abnormal globally on heat conduction in the
Yukawa chains under FK-modified on-site external potential while the local
temperature gradient appeared in the left region of system, see Fig.1(c).
The dependence of $J_{L}/J_{R}$, the heat flux ratio of opposite direction
for the Yukawa-Coulomb model, on parameter $\gamma _{2}$, the strength of
the FK-modified on-site external potential is plotted in Fig.3, where the
power law $J_{L}/J_{R}\propto 10^{0.24\gamma _{2}}$ is fitted in our
numerical parameter regime. Our numerical experiments shows the validity of
this power law until to $\gamma _{2}=4$ which indicates that the difference
of two directional heat current can possibly over one order. However in
practice of dusty plasmas it can only be realized by a relatively small $%
\gamma _{2}$ due to the test dust charge equals or less than the charge of
particles in chains. For example, for given $\chi =5$, $\gamma _{1}=1$ and $%
\beta =q_{t}/q\approx 1$ it corresponds to the practice ion flow Mach number 
$M=\sqrt{26}$ (usually in the sheath $M>1$), and $\gamma _{2}\approx 2$
which indicates $J_{L}/J_{R}\approx 3$.

\begin{figure}[tbp]
\hskip -2cm \vskip -0.75cm \includegraphics[width=0.50\textwidth]{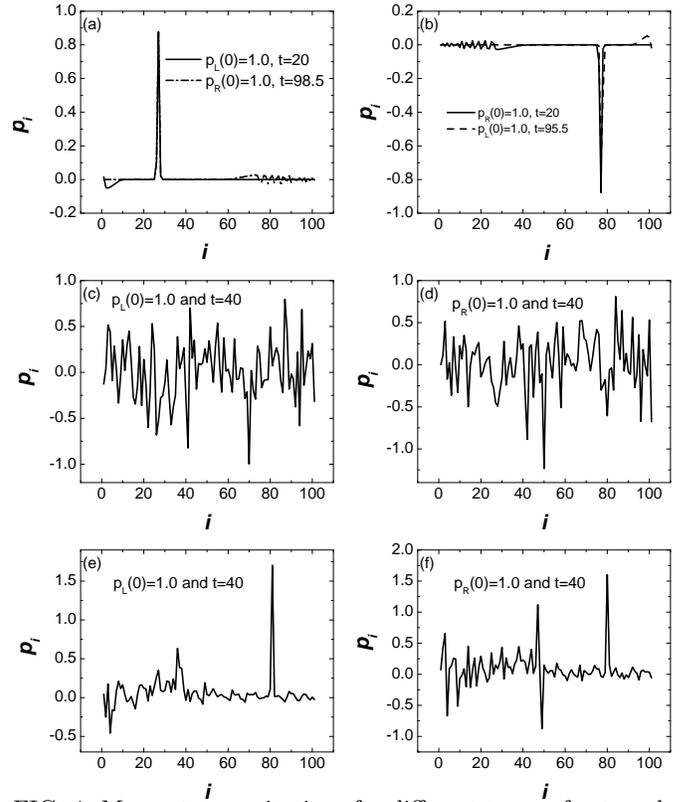} %
\vskip -2.25cm \vskip -2mm
\caption{Momentum excitations for different types of external potentials of
field-free [(a), (b)], FK [(c), (d)] and FK-modified [(e), (f)]. $p_{L}(0)$
and $p_{R}(0)$ denote the initial momentum kick on the left 1-th particle
and right 100-th particle, respectively. For (c)-(f) the parameters of $%
\protect\gamma _{1}=1$ and $\protect\gamma _{2}=2$ are given.}
\end{figure}

In order to see how the single kicked particle transfer energy to the
others, in Fig.4 we plotted the momentum excitations with different external
potentials of three cases of field-free, FK and FK-modified. We can see that
in the case of field-free there exists a solitary-like wave which can carry
energy independently and not transform it to others. In case of FK there
exist phonons which can transfer heat to the others. And in the case of
FK-modified there seems a combination of two kind waves mentioned above:
solitary-like and scatter phonons. Moreover it seems that there are always
more solitary-like waves which propagates in the left-direction than that in
the right direction. For example, there are three solitons in Fig.3(f) but
only one in Fig.3(e). Maybe this is one reason why in this case the
left-directional heat flux, $J_{L}$, is larger than that of right
directional, $J_{R}$.

\begin{figure}[tbp]
\vskip -0.5cm \includegraphics[width=0.62\textwidth]{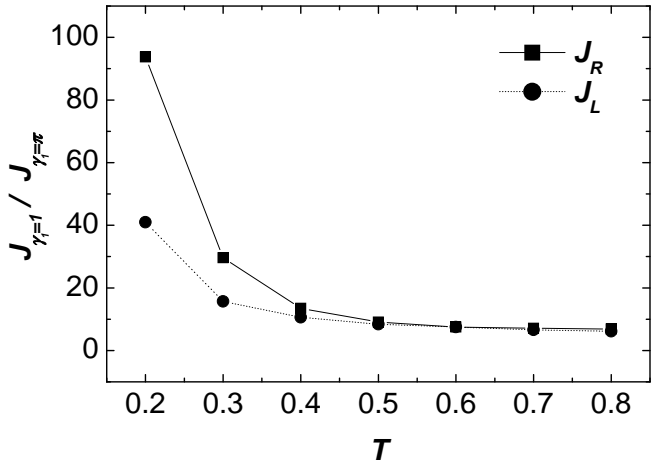} \vskip -9cm %
\vskip -2mm
\caption{Dependence of the heat flux ratio of two cases, for the Yukawa
chains with different inverse width of the FK-modified on-site external
potential, on temperature $T$. $\protect\gamma _{2}=2$ is given.}
\end{figure}

On the other hand according to the expression of heat flux, the ratio of
them of two-direction is approximately 
\begin{equation}
\frac{J_{L}}{J_{R}}\approx \sqrt{\frac{T_{+}}{T_{-}}}\frac{r_{>}}{r_{<}}\exp
\chi (r_{>}-r_{<})  \label{e8}
\end{equation}%
where $r_{>}$ and $r_{<}$ are the average inter-particle distance in the
region where the temperature is almost constant for $[T_{+},T_{-}]$,
right-direction heat-flow $J_{R}$, and $[T_{-},T_{+}]$, left-direction
heat-flow, respectively. In general it is shown that $r_{>}$ is a little
larger than $r_{<}$ by numerical experiments. It is not surprising that the
more phonons are scattered in the case of $[T_{+},T_{-}]$ than that of vise
versa. Therefore even if $r_{>}$ is just a very little greater than $r_{<}$, 
$J_{L}/J_{R}$ would be greatly larger than one. For example, $r_{>}\approx
1.08$ and $r_{<}\approx 0.92$, then we have $J_{L}/J_{R}\approx 3$. However
in the case of Harmonic interaction potential because of its symmetric
quadratic property about equilibrium position the ratio of opposite
directional heat flux is in order of $\sqrt{T_{+}/T_{-}}$, for example by
choosing $T_{+}=0.12$, $T_{-}=0.08$, the ratio of $J_{L}/J_{R}\approx 1.23$.
Our numerical results confirm this simple approximation analysis. Indeed if
we instead of harmonic interaction chains the heat flux difference for two
direction can not be over $23\%$ under asymmetric potential.

Finally we calculate heat flux in different potential width $\gamma
_{1}^{-1} $ (see Fig.5). It shows that we can also control the energy flow
by changing the potential width $\gamma _{1}^{-1}=\sqrt{M^{2}-1}/\chi $
through adjusting the ion flow velocity, which is associated with $M$ or and
the plasma particles density which is associated with $\chi$. Obviously the
great difference of heat flux at low temperature caused by the width of
potential smear out as the temperature increases to over $0.5$. It is noted
that the ratio of heat flux $J_{R}(\gamma_{1}=1)/ J_{R}(\gamma_{1}=\pi)$ is
about over two orders difference.

In summary we have studied the heat conduction problem in the system of 1D
Yukawa interaction chains with asymmetric on-site modified FK wake
potential. We can parametrically control the heat flux through this system
by changing the potential strength and width. Moreover it is also possible
to adjust the heat flux by changing the potential from FK to modified FK in
the Yukawa chains, thus we provide a simple method to change the properties
of the system, from a normal conductor obeying Fourier law, down to an
almost perfect insulator. Some theoretical consideration has also been
attempted to understand for our results and it is found that in the case of
Yukawa interaction potential, the asymmetric property about equilibrium
position has played a crucial role to improve the ratio of opposite
directional heat flux. It seems that there exist some modes which are
combinations of solitary-like and scattering phonons. Certainly more
formidable theoretical works is still challenging our research on this
problem in the future.

Authors would like to thank Drs. L. Wang, L. Yang and H. Zhang for their
useful discussions. This work was supported by the grants of the Hong Kong
Research Grants Council (RGC) and the Hong Kong Baptist University Faculty
Research Grant (FRG). B. S. X also acknowledge the support of funding in
part by the National Natural Science Foundation of China (NSFC) under Grant
Nos. 10275007 and 10135010.


\begin{references}

\bibitem{lep1} S. Lepri, R. Livi, and A. Politi, Phys. Rep. {\bf 377}, 1-80
(2003).

\bibitem{lep2} S. Lepri, R. Livi, and A. Politi, Phys. Rev. Lett. {\bf 78,}
1896 (1997).

\bibitem{hu1} B. Hu, B. Li, and H. Zhao, Phys. Rev. E {\bf 61}, 3828 (2000).

\bibitem{cas1} G. Casati, J. Ford, F. Vivaldi, and W. M. Visscher, Phys.
Rev. Lett. {\bf 52,} 1861 (1984).

\bibitem{pos1} H. A. Posch and Wm. G. Hoover, Phys. Rev. E {\bf 58,} 4344
(1998).

\bibitem{hu2} B. Hu, B. Li, and H. Zhao, Phys. Rev. E {\bf 57}, 2992 (1998).

\bibitem{sav1} A. V. Savin, and O. V. Gendelman, Phys. Rev. E {\bf 67},
041205 (2003).

\bibitem{alo1} D. Alonso, R. Artuso, G. Casati, and I. Guarneri, Phys. Rev.
Lett. {\bf 82,} 1859 (1999).

\bibitem{cas2} M. Terraneo, M. Peyrard and G. Casati, Phys. Rev. Lett. {\bf %
88,} 094302 (2002).

\bibitem{lin1} J. H. Chu and Lin I, Phys. Rev. Lett. {\bf 72,} 4009 (1994).

\bibitem{tho1} H. Thomas et al., Phys. Rev. Lett. {\bf 73,} 652 (1994).

\bibitem{vla1} S. V. Vladimirov and M. Nambu, Phys. Rev. E {\bf 52,} R2172
(1995).

\bibitem{xie1} B. S. Xie and M. Y. Yu, Phys. Rev. E {\bf 62,} 8501 (2000).

\bibitem{xie2} B. S. Xie and Z. A. Yang, Phys. Plasmas {\bf 9}, 4851 (2002).

\bibitem{sam1} D. Samsonov, A. V. Ivlev, R. A. Quinn and G. Morfill, Phys.
Rev. Lett. {\bf 88,} 095004 (2002).
\end{references}
\end{document}